# Directional Swimming of B. Subtilis Bacteria Near a Switchable Polar Surface

*Mahesha Kodithuwakku Arachchige, Zakaria Siddiquee, Hend Baza, Robert Twieg, Oleg D. Lavrentovich, Antal Jákli\**

Mahesha Kodithuwakku Arachchige, Zakaria Siddiquee, Hend Baza, Oleg D. Lavrentovich, Antal Jákli

Department of Physics, Kent State University, Kent, OH 44242, USA

E-mail: ajakli@kent.edu

Oleg D. Lavrentovich, Antal Jákli

Advanced Materials and Liquid Crystal Institute, Kent State University, Kent, OH 44242, USA

Robert Twieg

Department of Chemistry, Kent State University, Kent, OH 44242, USA

The dynamics of swimming bacteria depend on the properties of their habitat media. Recently it was shown that the motion of swimming bacteria dispersed directly in a non-toxic water-based lyotropic chromonic liquid crystal can be controlled by the director field of the liquid crystal. Here we investigate whether the macroscopic polar order of a ferroelectric nematic liquid crystal ($N_F$) can be recognized by bacteria B. Subtilis swimming in a water dispersion adjacent to a glassy $N_F$ film by surface interactions alone. We show that B. Subtilis tends to move in the direction antiparallel to the spontaneous electric polarization at the $N_F$ surface. Their speed was found to be the same with or without a polar $N_F$ layer. In contrast to observation on crystal ferroelectric films, the bacteria do not get immobilized. These observations may offer a pathway to creation of polar microinserts to direct bacterial motion in-vivo.

# 1. Introduction

The dynamics of swimming bacteria depend on the bulk properties of their habitat media such as the viscosity[1] and on the structure of interfaces.[2] Concerning the effect of bulk properties, recently swimming bacteria have been introduced as a part of synthetic active systems, by dispersing them directly in a non-toxic water-based lyotropic chromonic liquid crystal (LCLC)[3–15] or forming active aqueous droplets in a thermotropic liquid crystal.[16,17] The presence of the passive liquid crystal component allows one to control the individual[7–11,18] and collective[13–15] dynamics of the bacteria. For example, when the concentration of bacteria is low, the director of an LCLC defines the direction of swimming.[7–10,13,18] In non-polar liquid crystal media such as the LCLCs, the axis of orientational order is apolar, thus in a monodomain sample with a uniform director, the individual swimming of bacteria cannot be rectified into a polar flow. The polarity of swimming direction could be triggered only by the splay and bend deformations of the LCLC director field, which are themselves polar entities.[12–16] Directional swimming of bacteria in uniform liquid crystals can be envisioned if the bacteria are placed in a polar lyotropic liquid crystal. Although a lyotropic analog of a thermotropic ferroelectric phase has been suggested,[19] its polar nature has not been shown. Ferroelectric liquid crystals with switchable macroscopic polarization may have one-,[20] two-[21–25] and three-dimensional fluid order.[26–29] They are thermotropic and hydrophobic, so they do not mix with the aqueous environment of the bacteria, but they can be used as an adjacent medium to a bacterial dispersion with a sharp interface. In fact, bacteria are known to swim closely to interfaces, both in isotropic[1,2] and anisotropic LCLC environment,[9–11] thus one may assume that surface forces will have an effect on their movement.

The interface effect of ferroelectric crystals such as $LiNbO_3$ (Lithium Niobate) and $LiTaO_3$ (Lithium Tantalate) have already been tested for some bacteria. It was found that the evanescent field of the spontaneous polarization of these ferroelectric crystals can trap [30] and/or orient bacteria along the field.[31]

In this paper we show that the macroscopic polar order of a vitrified ferroelectric nematic ($N_F$) liquid crystal (LC) material RT11165 [32] (Figure 1(a)) can be recognized by bacteria B. Subtilis swimming in a water dispersion adjacent to the glassy $N_F$ film. We chose the newly discovered $N_F$ liquid crystals, because they have the largest spontaneous polarization with the lowest threshold voltage for switching among ferroelectric liquid crystals. We have found that fluid nematic, smectic and columnar ferroelectric liquid crystals tend to flow away under the weight of the aqueous material layered on top of them. Therefore, the room temperature glassy nature of the studied $N_F$ material with a polarization that can be switched above room temperature is essential to form a stable flat interface between the liquid crystal film and the aqueous bacteria droplet. The geometry of the sample with the definition of the directions and the sample preparation process are shown in Figure 1(b-d).

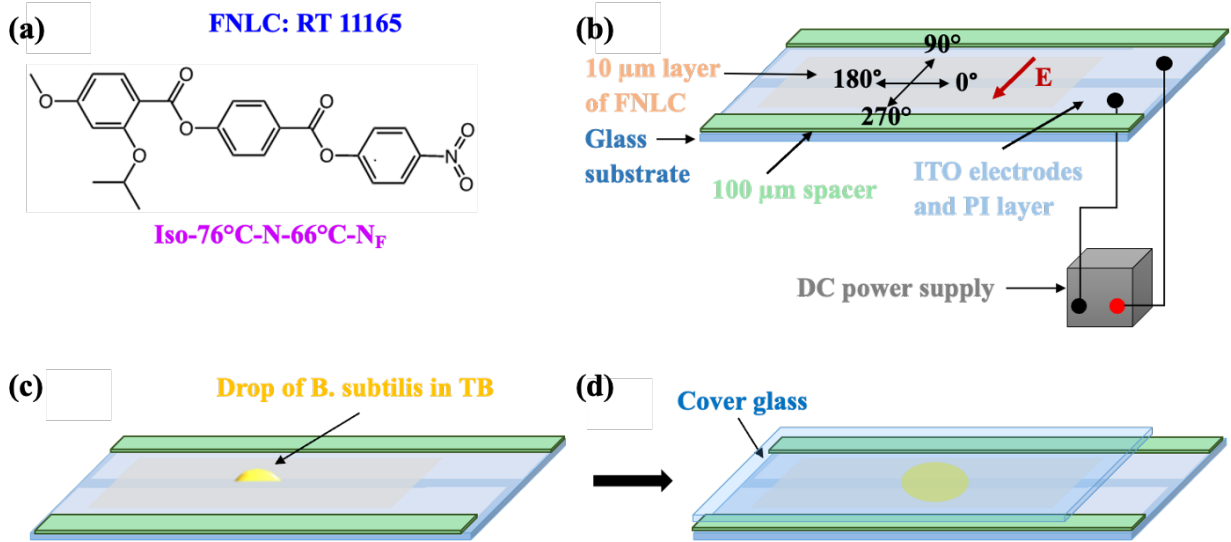

**Figure 1:** The molecular structure of RT 11165, the geometry of the sample with the definition of the directions and illustration of the sample preparation process. (a) Molecular structure and phase sequence of the ferroelectric nematic liquid crystal, RT 11165. (b) Schematics of the glass substrate covered with ITO electrodes separated by 0.5 mm gap to apply a horizontal DC electric field $\vec{E}$ along 90° and 270° directions. The substrates are coated by about 10 μm thick RT11165 film. The substrates also contain 100 μm thick spacers at their edges (shown by green strips). (c) A drop of Terrific Broth (TB) containing B. Subtilis is added after removing the electric field. (d) A cover glass is placed on the spacers thus providing a 100 μm thick film of bacteria dispersed in TB.

## 2. Experimental Results

On cooling from the isotropic liquid phase RT 11165 transitions to the conventional non-polar nematic (N) phase at 76 °C, then at 66 °C to the ferroelectric nematic ($N_F$) phase, which vitrifies at ∼20 °C.[32] Although in the $N_F$ phase the ferroelectric polarization could be measured only down to 50 °C, where $P_o \approx 6.5\ \mu Ccm^{-2}$, it could be fully switched even at 30 °C by 0.1 V$\mu m^{-1}$ DC field in about 10 s corresponding to a rotational viscosity of $\gamma_1 \sim 10^5$ Pa.[32] Therefore the polarization of the material can be aligned by an external electric field prior to adding the aqueous droplet of bacteria on the top. In accordance with other studies,[33,34] the direction of the polarization of thin RT11165 films could also be set by a rubbed underlying polyimide (PI 2555) layer so that the polarization direction is opposite to the rubbing direction. Polarizing optical microscopy (POM) images of the $N_F$ textures are shown in **Figure 2**a-f on differently treated substrates after removing the electric field and just before adding the bacteria. The POM textures are dark, indicating nearly perfect LC alignment in **Figure 2**a-e when the $N_F$ is aligned by either rubbed surface or electric field, or both. The texture is bright and inhomogeneous when there is neither rubbing nor electric field alignment (Figure 2f).

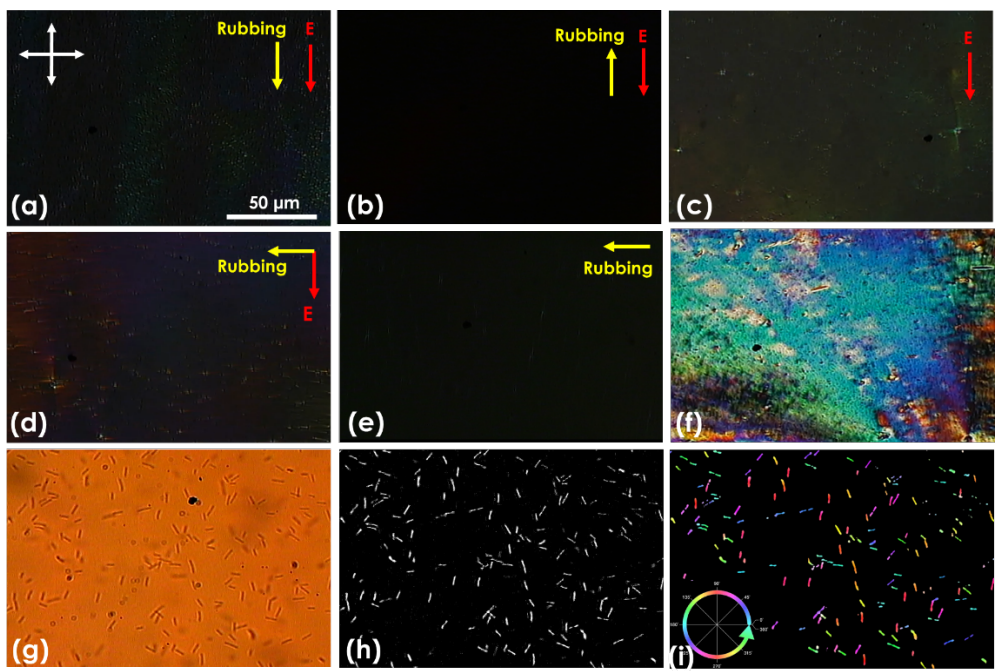

**Figure 2:** POM textures after different alignment procedures (a-f) and snapshots of microscopic images of bacteria (g-i). (a) Rubbing and electric field are parallel. (b) Rubbing and electric field are anti-parallel. (c) Electric field with no alignment layer. (d) Rubbing and electric field are perpendicular. (e) Rubbing with no electric field. (f) No rubbing and no electric field. (g) A snapshot of bacteria from the original video. (h) Binary image produced from the original snapshot. (i) Color coded image according to the orientation of bacteria.

Figure 2g shows a snapshot of a typical microscopy image without any processing. To improve the contrast, the image was inverted, the background was removed and then converted into a binary image as shown in Figure 2h. These binary image stacks were used for Particle Tracking Velocimetry (PTV) and Particle Image Velocimetry (PIV) analysis. PTV analysis tracks the path of individual bacteria and in each frame (~10 frame/s) outputs the track counts in each direction from 0° to 360°. PIV analysis outputs the total number of bacteria swimming in the same direction from 0° to 360°. The Binary images were then color coded according to the orientation of bacteria as shown in Figure 2i for Orientation J analysis. Orientation J adds colors to images of individual bacteria depending on the direction they are oriented at a given moment and outputs the distribution of orientation of bacteria in a range of angular directions from -90° to 90° as shown in Figure S1 of the Supporting Information (SI).

We recorded the motion of the bacteria in two distance ranges from the substrate: range 1, labeled as $h = 0$ is between 0 and 10 μm from the substrate, and range 2, labeled as $h = 10$, is between 10 μm and 20 μm from the substrate. All data were normalized by dividing the number of bacteria or their tracks in each direction by the total number of bacteria or their tracks and plotted with respect to the angle between 0° and 360°. In PIV and PTV plots, the original data points were smoothed with a sliding average of 11 data points. Methods of analysis are described in detail in SI.

PTV results in the $h = 0$ range are shown in **Figure 3**. To facilitate the discussion, we refer to the axis along the channel between the two electrodes as 0° direction as shown in **Figure 1**b. The dc field can be applied along the 90° or 270° direction.

When there is no underlying $N_F$ layer, the normalized bacteria track scatters significantly as seen in **Figure 3**a, and within the error of the measurement, no clear tendency can be seen neither in the original nor in the smoothed track count. As the blue data points in **Figure 3**b show, there is also a large scattering of the track count when there is an $N_F$ layer that has neither been aligned by an electric field nor by rubbing of the PI. This indicates that the smearing direction of the $N_F$ film during deposition does not have a significant effect on the alignment of the $N_F$. This is understandable as the smearing happened in the isotropic phase, so it does not have any effect on the alignment in the $N_F$ phase.

The peaks in cells with aligned $N_F$ layers are clearly much stronger than the measuring error, as seen in the red data in **Figure 3**b when the PI coating has been rubbed along 180°, and in the smoothed curves in **Figure 3**c and Figure 3d where electric fields were applied on the $N_F$ films.

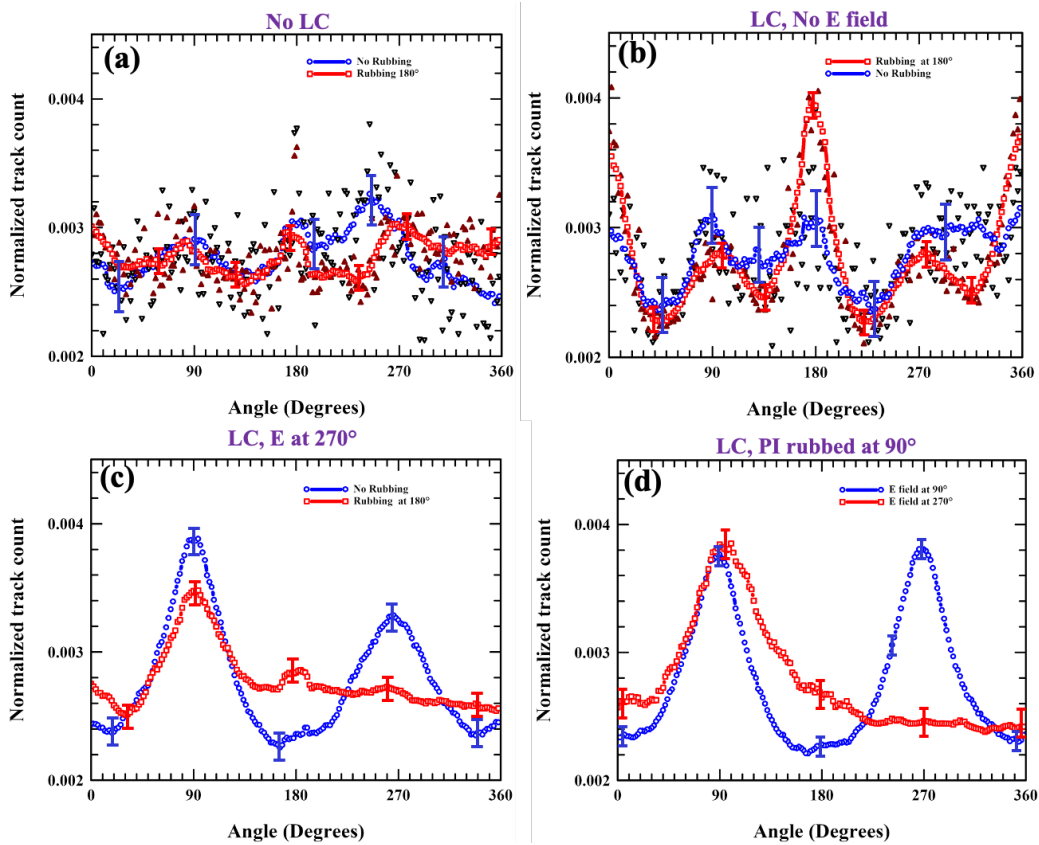

**Figure 3:** PTV results at $h$=0. Red and blue triangles in (a) and (b) are original data and blue circles and red squares are smoothed data points. (a) Normalized track count when there is no underlying $N_F$ film. (b) Normalized track count when there is an $N_F$ film without an electric field applied. Blue data show the case when there is no rubbing of the underlying PI film and red data show the results when the LC film is aligned along 180° by rubbed PI. (c) Normalized track count when the LC film is aligned only by applying electric field in 270°. Blue data: no PI rubbing; red data: PI rubbed along 180°. (d) Results for underlying PI rubbed along 90° while electric field is applied in the same direction (blue data) and in the opposite direction (red data). Half-lengths of the error bars have been calculated from the average deviation of the original data from the smoothed data. The angle conventions are shown in Figure 1b.

When only rubbing aligns the $N_F$ film along 180° (red data in **Figure 3**b), the additional peak along 0° indicates that the rubbing does not fully align the polarization. Similarly, when the $N_F$ film was aligned without a rubbing by only an electric field applied along 270° (blue points in **Figure 3**c), two peaks are seen: the larger is along 90°, i.e., opposite to the field direction, and the smaller is along the electric field. This again indicates that the electric field alone is not sufficient to align all polarization.

Red data points of Figure 3c show that when the rubbing direction (180°) was perpendicular to the electric field (270°), there is one large peak at 90° and two small peaks along the field direction and along the rubbing direction. In this case, the background values are relatively large, indicating less perfect alignment in accordance with Figure 2d.

Blue data points of Figure 3d show results when the rubbing and the electric field are in the same 90° direction. It reveals two equally strong peaks at 90° (along rubbing and field) and at 270° (opposite to electric field rubbing). This is because, while the electric field aligns the polarization along the field, the rubbing aligns the polarization in the opposite direction [33,34]. After field removal the polarization direction slowly (in hours) randomizes if the surface alignment does not support the field-induced alignment. Although we apply bacteria within 10 minutes after removing the electric field, the field induced alignment slightly deteriorates without sufficient surface alignment by the time of the measurements.

Using a substrate rubbed opposite to the electric field direction halts the relaxation after the removal of the electric field. This is verified by the red data points in Figure 3d that show the results when the directions of the rubbing (90°) and the electric field (270°) are opposite. Now there is only one strong peak at 90°, i.e., along the rubbing and opposite to the electric field. This is because now both the electric field and the rubbing align the polarization along 270°. Therefore, we conclude that the bacteria tend to swim opposite to the direction of the ferroelectric polarization of the $N_F$ film.

PTV results at $h$=10 in Figure S2 of the SI, show similar tendencies with somewhat weaker correlations. This is expected, as the bacteria are further away from the surface, thus less influenced by the direction of the polarization. We also noted that there are fewer bacteria in the $h$=10 range, which is observed for other pusher type bacteria that tend to aggregate near solid walls.[2]

PIV and Orientation J results shown in figures S3 - S6 in the SI also agree with the PTV results, thus reinforcing our conclusion that the bacteria tend to move opposite to the direction of the ferroelectric polarization. We note that the Orientation J results specify only the apolar alignment of the bacteria and do not distinguish 0° from 180° and 90° from 270°. The agreement between the PTV and Orientation J results with these constraints demonstrates that the bacteria are swimming along their long axis.

Helical swimmers such as B. Subtilis typically have circular motion in a plane parallel to an interface.[1,2] The sence of circulation is often clockwise (CW) near a solid substrate and counterclockwise (CCW) near a free surface. The sence of circulation is determined by complex hydrodynamic interactions that depend on the slip condition, distance to the interface, shape of the bacterium, etc.[1,2] Indeed, we observed that the paths taken by B. Subtilis bacteria were not consistently straight but frequently showed nearly circular motion with ~85% CCW and 15% CW rotation. This prompted the investigation of the ratio of linear to elliptical trajectories, where trajectories with an arclength of 80 μm and radius of curvature larger than 126 μm are considered quasi rectilinear, while trajectories of an arclength of 80 μm and a radius of curvature smaller than 126 μm are considered as quasi circular. To accomplish this, positional data for all bacteria visible in each video was extracted using the TrackMate plugin in ImageJ, providing coordinates for each bacterium across all time frames, as illustrated in Figure S7-S9. The trajectory statistical results are summarized in Figure S10. It was found that ratio of the linear to elliptical track counts is $2.9 \pm 0.2$ and $2.4 \pm 0.2$ at $h=10$ and $h=0$, respectively for a substrate that contains no LC and there was no electric field applied before deposition of the bacteria. This shows an increasing tendency for circular motion closer to the substrate. These numbers did not change significantly ($2.7 \pm 0.2$ and $2.4 \pm 0.2$ at $h=10$ and $h=0$) when the glass substrate (still with no LC) was rubbed, indicating negligible effect of the surface rubbing on the circular motion. In contrast, the presence of the $N_F$ layer had a very significant effect on the ratio of the quasi rectilinear and quasi circular motion, as it decreased to a range between $0.7 \pm 0.1$ and $1.3 \pm 0.1$ at $h = 0$ and $1.1 \pm 0.1$ and $1.7 \pm 0.15$ at $h = 10$. The ratio of the CCW to CW circulation was still 85:15. The ratio of minor to major axes (circularity) of the ellipses was found to be close to 1, about $0.9 \pm 0.1$ for $h=10$ and $h=0$, independently of the presence of the LC layer and other additional treatment (rubbing or electric field). It means that the elliptical trajectories do not bias much the ratio of the linear to circular trajectories. The increase of the fraction of quasi circular motion near the $N_F$ substrate indicates a potential attraction of the bacteria to the surface.[1,2,35]

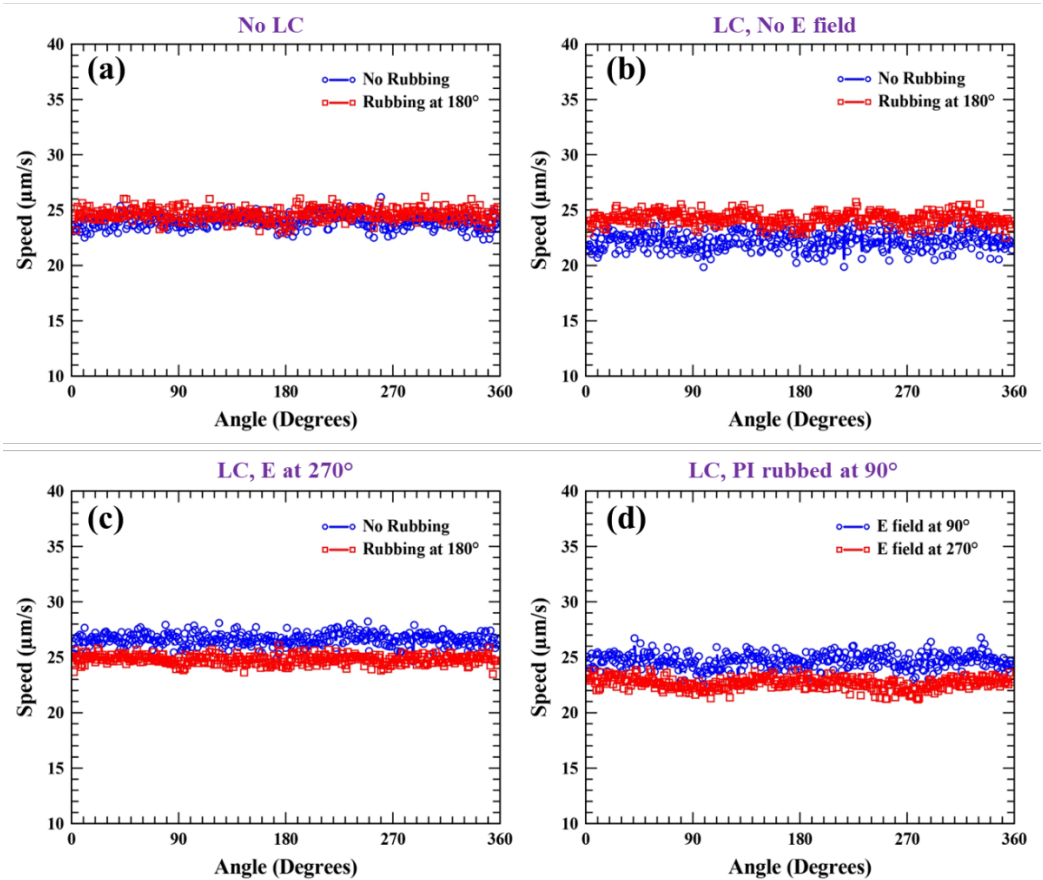

**Figure 4:** Distribution of average bacterial speed in different orientations. (a) Speed distribution when there is no underlying $N_F$ film. (b) Speed distribution when there is an $N_F$ film without an electric field applied. Blue data show the case when there is no rubbing of the underlying PI film and red data show the results when the LC film is aligned along 180° by rubbed PI. (c) Speed distribution when the LC film is aligned only by applying electric field in 270°. Blue data: no PI rubbing; red data: PI rubbed along 180°. (d) Results for underlying PI rubbed along 90° while electric field is applied in the same direction (blue data) and in the opposite direction (red data).

Distribution of bacterial speed in different orientations, for 8 different geometries studied is shown in **Figure 4**. Average bacterial speed over each degree is plotted with respect to azimuthal angle of the orientation. The bacteria speed appears to be unaffected by the orientation. For the situation when the electric field is applied at 270° opposite to the rubbing direction (90°), i.e., when both the electric field and the rubbing aligns the polarization along 270°, there is a slight pattern indicating slower motion along the polarization than perpendicular to it. Further analysis indicates that this angular dependence is resulting only from bacteria that are moving in a straight line and not the elliptical ones. Slight variation of bacteria speed in different configurations may be related to the health of bacteria and available nutrients in the medium.

## 3. Discussion

In this study we found (see Figure 3d) that B. Subtilis bacteria dispersed in Terrific Broth are getting oriented and swim with ~40% higher probability in the direction opposite to the polarization $\vec{P}$ of an underlying glassy ferroelectric nematic film than in any random direction. Such a tendency is schematically illustrated in Figure 5. In contrast to observation on crystal ferroelectric substrates,[30,31] the bacteria did not get immobilized and within the error of the measurements, their speed was the same as without a polar $N_F$ layer (see Figure 4).

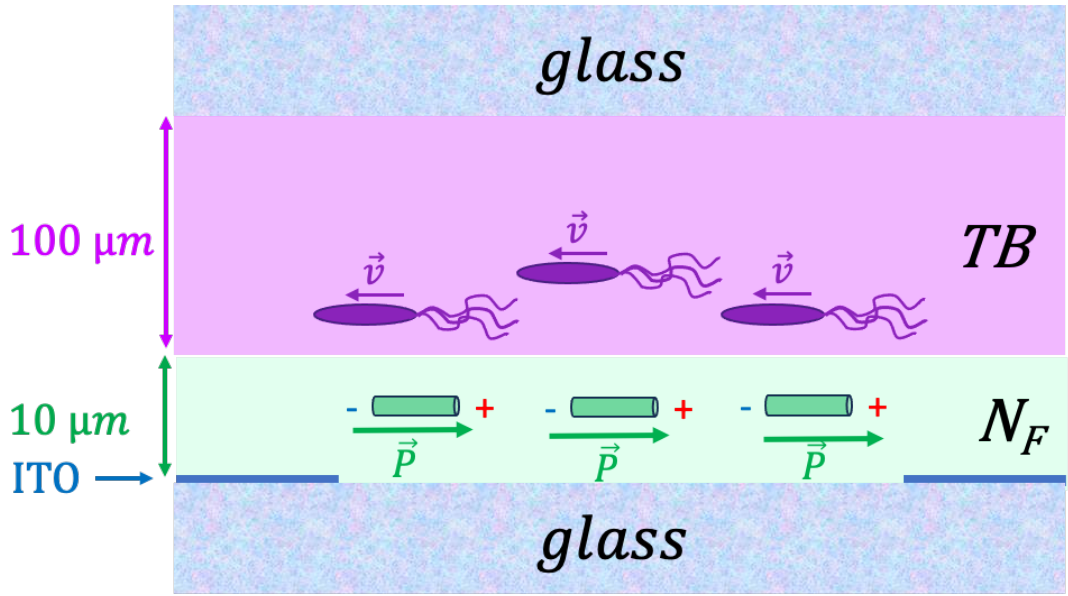

**Figure 5:** Illustration of B. Subtilis bacteria motion opposite to the polarization direction of an underlying ferroelectric nematic liquid crystal with uniform polarization $\vec{P}$. Note: in reality, the size of ferroelectric nematic liquid crystal molecules is 3-4 orders of magnitude smaller than that of the bacteria.

Several studies have shown that the surface of bacteria is highly charged due to the presence of various structures extending away from the membrane.[36–39] At near-neutral pH values, both gram-positive and gram-negative bacteria have a net negative charge on their surface.[39] Also, they have negative zeta potential and among them B. Subtilis has the highest magnitude.[40] It was also shown that B. Subtilis has an asymmetric charge distribution across the cell wall with the regions close to polar caps being more electronegative.[38,39,41–44]

In case of our glassy $N_F$ LC [32] the polarization field ends at an angle $\alpha$ with respect to the plane on the insulating PI layer above the in-plane electrodes, thus leading to a depolarization field directed perpendicularly to the film, of the magnitude $E_{dep} = -\frac{P\sin\alpha}{\varepsilon_o\varepsilon}$. Such a field will attract free ions above the edge of the electrodes thus attracting potassium ions to the edge where $E_{dep} < 0$. The range of the gradient, however, is likely to be negligible compared to the 0.5 mm gap between the electrodes, thus it is unlikely that it leads to a movement called electrotaxis.[43,44]

The orientation of the bacteria placed in the evanescent field $\vec{E}$ of an underlying spatially periodic ferroelectric polarization of Fe: $LiNbO_3$ was explained by a torque $\vec{\Gamma} = \vec{p} \times \vec{E}$ of the induced dipole moment $\vec{p}$ acting on a cylindrically shaped bacterium. The trapping was explained by a force $\vec{F} \propto \vec{\nabla}\vec{E}$ due to the gradient of the evanescent field. However, the depolarization field-induced gradient of ionic concentration is expected by to screened over short distances on the order of a sub-micrometer Debye screening length, thus it is not clear how these limited regions in our case can bias the trajectories monitored in the middle of the 0.5 mm gap.

Another possibility is that the bacteria can sense the direction of polarization at the substrate electrostatically. B. Subtilis are microswimmers of a pusher type; their swimming is accompanied by two water jets directed away from the bacterium along the rod-like body and attraction of the fluid along the equatorial plane. Because of this flow pattern, the surface (i) orients the rod-like bacterium parallel to itself, (ii) imposes an attractive force on the bacterium, and (iii) contributes a circular trajectory.[1,2] The forces responsible for circular trajectories decrease as the distance to the substrate increases.[1,2] The observed polar bias in bacterial trajectories could be associated with the polar nature of the $N_F$ substrate. Any rubbed or field-poled surface of an $N_F$ should feature an in-plane direction of polarization.[33,34] If the bacterium exhibits a permanent longitudinal dipole directed from the flagellum towards the head, its antiparallel arrangement with the $N_F$ polarization should cause electrostatic attraction, while a parallel arrangement would imply repulsion. One can thus expect that the bacteria swimming antiparallel to $\vec{P}$ would spend more time near the surface as compared to the bacterium swimming in an opposite direction, which explains the observed polar bias. The possibility of permanent dipoles on the bacterial surface was supported by the experiments on motile flagellated E. Coli swimming in an externally applied dc electric field.[45] The dipoles can also be induced by interactions with the $N_F$ substrate, with would result in a similar electrostatic attraction. Although we do not know the precise mechanisms of electrostatic attraction, its presence is supported by the observation that the ratio of rectilinear to circular trajectories decreases dramatically in the presence of the $N_F$ substrate. As demonstrated by independent studies, circular motion is more likely to occur when the distance of a bacterium to the surface decreases.[1,2,35]

## 4. Future prospects

Directed motion of bacteria on polar liquid crystal surfaces may have several advantages. (1) With the use of patterned electrodes the direction various shape of the polarization field can be achieved and can also be reconfigured after the previous bacteria and TB washed out for multiple uses. (2) With suitable polarization pattern, the bacteria's density can be controlled for special observations. (3) The simplicity and cost-effective procedure involving $N_F$ layeres enables their use in biochemical and clinical laboratories without expensive equipments. (4) The cation-free interaction makes $N_F$ film directed swimming environmentally friendly tool. (5) As the $N_F$ film is not dissolving in aqueous environment, one may create polar microinserts to direct bacterial motion in-vivo.

## 5. Experimental section

### *5.1 Bacteria dispersion preparation*

The bacteria are initially grown on a lysogeny broth (LB, purchased from Teknova) agar plate, then transferred to a tube containing terrific broth (TB, purchased from Sigma Aldrich) liquid medium and grown in shaking incubator at 35 °C for 7-9 hours until the maximum saturation of bacterial concentration is reached. The liquid medium containing bacteria is then extracted from the tube and centrifuged for 30 seconds at 4000 rpm to filter out the less active bacteria. The concentration of the resulting bacteria solution is $4 \times 10^{13}$ m$^{-3}$.

*5.2 Substrate preparation*

A few nm thick indium tin oxide (ITO) layer sputtered on glass substrates were sonicated in detergent (Cavi-clean ultrasonic detergent) water at 60 °C for 15 minutes and then washed with distilled water and isopropyl alcohol (IPA) respectively, followed by drying at 90 °C for 15 minutes. After that they were etched with 0.5 mm gap in the middle (**Figure 1**b) by photolithography. Substrates with electrodes were cleaned again according to the above-mentioned procedure. Cleaned substrates were then spin-coated by about 50 nm thick PI 2555 (dissolved in T9039 thinner) polyimide layer and rubbed unidirectionally with a rubbing block covered with a velvet cloth. Control measurements were also carried out for the bare glass-ITO substrates without PI 2555 and the substrates with an unrubbed layer of PI2555.

*5.3 Method*

To make ~10 μm thick RT11165 films, a drop was placed on such substrates, heated to 80 °C and smeared unidirectionally with a needle. Then two layers of 50 μm thick cello tapes were placed along each of the two long sides of the glass substrate to make two 100 μm thick spacers. Two wires were soldered to the ITO electrodes and connected to the DC power supply. A 0.1 Vμm$^{-1}$ DC electric field is used to align the $N_F$ polarization parallel to the field in the gap area. After turning off the field, the director alignment (and consequently the polarization field) remained largely unchanged for over 30 minutes. A drop of terrific broth (TB, purchased from Sigma Aldrich) containing B. Subtilis bacteria (wild-type strain 1085, body length of 5-7 μm) at $4 \times 10^{13}$ m$^{-3}$ concentration was placed on each of the differently treated substrates (**Figure 1**c). Finally, a cover glass is placed on 100 μm thick spacers to make a cell and to confine the bacteria, Figure 1d. Experiments were performed on different substrates with eight different combinations of rubbing and electric field directions, with and without an $N_F$ film.

*5.4 Analyzing Motion of Active Matter in 2D*

To determine the orientational distribution of the movement of the B. Subtilis swimmers, we employed both Particle Image Velocimetry (PIV) and Particle Tracking Velocimetry (PTV). PTV analysis tracks the path of individual bacteria and in each frame (~10 frame/s) outputs the track counts in each direction from 0° to 360°. PIV analysis outputs the total number of bacteria swimming in the same direction from 0° to 360°. The PIV and PTV analyses are three-step procedures. First, the video footage was converted into grayscale image stacks. Second, ImageJ was utilized to pre-process the image stacks. HWada TPIV (ImageJ plugin) was employed for PIV analysis, while the TrackMate (ImageJ plugin) was utilized for PTV analysis. In the last step, the output files of the PIV (Positions, Magnitude and Angles for pixel shift in each interrogation window of the image stacks) and PTV (Bacteria label, Positions in each frame of the image stacks) obtained from the ImageJ plugins were imported into Python 3.10, where the distribution was calculated and visualized by plotting figures. The image pre-processing, HWada TPIV and TRackMate Data export, and Python PIV, PTV and Orientation JAnalyses are described in SI.

## *6.* Supporting Information

Supporting Information is available from the Wiley Online Library or from the author.

## 7. Acknowledgements

This work was supported by the US National Science Foundation under grant DMR-2210083 and DMR-2215191 and ECCS-2122399.

## 8. Conflict of Interest Statement

The authors declare no conflict of interest.

## 9. Data Availability Statement

The data that support the findings of this study are available from the corresponding author upon reasonable request.

# 10. References

[1]  E. Lauga, *Annu Rev Fluid Mech* **2016**, *48*, 105.

[2]  D. Lopez, E. Lauga, *Physics of Fluids* **2014**, *26*, 071902.

[3]  A. Kumar, T. Galstian, S. K. Pattanayek, S. Rainville, *Molecular Crystals and Liquid Crystals* **2013**, *574*, 33.

[4]  S. Zhou, *Liquid Crystals Today* **2018**, *27*, 91.

[5]  Y. K. Kim, X. Wang, P. Mondkar, E. Bukusoglu, N. L. Abbott, *Nature* **2018**, *557*, 539.

[6]  M. M. Genkin, A. Sokolov, O. D. Lavrentovich, I. S. Aranson, *Phys Rev X* **2017**, *7*, 011029.

[7]  S. Zhou, A. Sokolov, O. D. Lavrentovich, I. S. Aranson, *Proc Natl Acad Sci U S A* **2014**, *111*, 1265.

[8]  P. C. Mushenheim, R. R. Trivedi, H. H. Tuson, D. B. Weibel, N. L. Abbott, *Soft Matter* **2014**, *10*, 88.

[9]  P. C. Mushenheim, R. R. Trivedi, S. S. Roy, M. S. Arnold, D. B. Weibel, N. L. Abbott, *Soft Matter* **2015**, *11*, 6821.

[10] P. C. Mushenheim, R. R. Trivedi, D. B. Weibel, N. L. Abbott, *Biophys J* **2014**, *107*, 255.

[11] S. Zhou, O. Tovkach, D. Golovaty, A. Sokolov, I. S. Aranson, O. D. Lavrentovich, *New J Phys* **2017**, *19*, 055006.

[12] O. D. Lavrentovich, *Liq Cryst Rev* **2020**, *8*, 59.

[13] R. Koizumi, T. Turiv, M. M. Genkin, R. J. Lastowski, H. Yu, I. Chaganava, Q. H. Wei, I. S. Aranson, O. D. Lavrentovich, *Phys Rev Res* **2020**, *2*, 033060.

[14] T. Turiv, R. Koizumi, K. Thijssen, M. M. Genkin, H. Yu, C. Peng, Q. H. Wei, J. M. Yeomans, I. S. Aranson, A. Doostmohammadi, O. D. Lavrentovich, *Nat Phys* **2020**, *16*, 481.

[15] C. Peng, T. Turiv, Y. Guo, Q.-H. Wei, O. D. Lavrentovich, *Science (1979)* **2016**, *354*, 882.

[16] M. Rajabi, H. Baza, T. Turiv, O. D. Lavrentovich, *Nat Phys* **2021**, *17*, 260.

[17] M. Rajabi, H. Baza, H. Wang, O. D. Lavrentovich, *Front Phys* **2021**, *9*, 752994.

[18] A. Sokolov, S. Zhou, O. D. Lavrentovich, I. S. Aranson, *Phys Rev E Stat Nonlin Soft Matter Phys* **2015**, *91*, 013009.

[19] J. R. Bruckner, F. Giesselmann, *Crystals (Basel)* **2019**, *9*, 568.

[20] H. Bock, W. Helfrich, *Liq Cryst* **1992**, *12*, 697.

[21] S. T. Lagerwall, *Ferroelectric and Antiferroelectric Liquid Crystals*, Wiley, **1999**.

[22] A. Jakli, A. Saupe, *One- and Two-Dimensional Fluids*, CRC Press, Taylor And Francis, **2006**.

[23] R. B. Meyer, L. Liebert, L. Strzelecki, P. Keller, *Journal de Physique Letters* **1975**, *36*, 69.

[24] H. Takezoe, J. Watanabe, *Molecular Crystals and Liquid Crystals Science and Technology Section A: Molecular Crystals and Liquid Crystals* **1999**, *328*, 325.

[25] T. Niori, T. Sekine, J. Watanabe, T. Furukawa, H. Takezoe, *J Mater Chem* **1996**, *6*, 1231.

[26] R. J. Mandle, S. J. Cowling, J. W. Goodby, *Physical Chemistry Chemical Physics* **2017**, *19*, 11429.

[27] H. Nishikawa, K. Shiroshita, H. Higuchi, Y. Okumura, Y. Haseba, S. I. Yamamoto, K. Sago, H. Kikuchi, *Advanced Materials* **2017**, *29*, 1702354.

[28] O. D. Lavrentovich, *Proc Natl Acad Sci U S A* **2020**, *117*, 14629.

[29] X. Chen, E. Korblova, D. Dong, X. Wei, R. Shao, L. Radzihovsky, M. A. Glaser, J. E. MacLennan, D. Bedrov, D. M. Walba, N. A. Clark, *Proc Natl Acad Sci U S A* **2020**, *117*, 14021.

[30] O. Gennari, V. Marchesano, R. Rega, L. Mecozzi, F. Nazzaro, F. Fratianni, R. Coppola, L. Masucci, E. Mazzon, A. Bramanti, P. Ferraro, S. Grilli, *ACS Appl Mater Interfaces* **2018**, *10*, 15467.

[31] L. Miccio, V. Marchesano, M. Mugnano, S. Grilli, P. Ferraro, *Opt Lasers Eng* **2016**, *76*, 34.

[32] A. Adaka, P. Guragain, K. Perera, P. Nepal, B. Almatani, S. Sprunt, J. Gleeson, R. J. Twieg, A. Jákli, *Liq Cryst* **2024**, *51*, 1140.

[33] B. Basnet, M. Rajabi, H. Wang, P. Kumari, K. Thapa, S. Paul, M. O. Lavrentovich, O. D. Lavrentovich, *Nat Commun* **2022**, *13*, 3932.

[34] H. Kamifuji, K. Nakajima, Y. Tsukamoto, M. Ozaki, H. Kikuchi, *Applied Physics Express* **2023**, *16(7)*, 071003.

[35] G. Li, L.-K. Tam, J. X. Tang, *PNAS* **2008**, *105*, 18355.

[36] A. Revil, E. Atekwana, C. Zhang, A. Jardani, S. Smith, *Water Resour Res* **2012**, *48*, W09545.

[37] W. W. Wilson, M. M. Wade, S. C. Holman, F. R. Champlin, *J Microbiol Methods* **2001**, *43*, 153.

[38] T. J. Beveridge, L. L. Graham, *Microbiol Rev* **1991**, *55*, 684.

[39] E. M. Sonnenfeld, T. J. Beveridge, A. L. Koch, R. J. Doyle, *J Bacteriol* **1985**, *163*, 1167.

[40] W. Pajerski, D. Ochonska, M. Brzychczy-Wloch, P. Indyka, M. Jarosz, M. Golda-Cepa, Z. Sojka, A. Kotarba, *Journal of Nanoparticle Research* **2019**, *21*, 186.

[41] E. M. Sonnenfeld, T. J. Beveridge, R. J. Doyle, *Can J Microbiol* **1985**, *31*, 875.

[42] J. Humphries, L. Xiong, J. Liu, A. Prindle, F. Yuan, H. A. Arjes, L. Tsimring, G. M. Süel, *Cell* **2017**, *168*, 200.

[43] P. Chong, B. Erable, A. Bergel, *Biofilm* **2021**, *3*, 100048.

[44] B. Cortese, I. E. Palamà, S. D'Amone, G. Gigli, *Integr. Biol.* **2014**, *6*, 817.

[45] W. Shi, B. A. D. Stocker, J. Adler, *J Bacteriol* **1996**, *178*, 1113.

# Supporting Information

**Directional Swimming of B. Subtilis Bacteria Near a Switchable Polar Surface**

*Mahesha Kodithuwakku Arachchige, Zakaria Siddiquee, Hend Baza, Robert Twieg, Oleg D. Lavrentovich, Antal Jákli\**

1. **Analyzing Motion of Active Matter in 2D**

HWada TPIV (ImageJ plugin) was employed for PIV analysis, while the TrackMate (ImageJ plugin) was utilized for PTV analysis. In the last step, the output files of the PIV (Positions, Magnitude and Angles for pixel shift in each interrogation window of the image stacks) and PTV (Bacteria label, Positions in each frame of the image stacks) obtained from the ImageJ plugins were imported into Python 3.10, where the distribution was calculated and visualized by plotting figures.

**1.1 Image Pre-Processing**

Prior to analyzing the distribution of directional orientation of active swimmers, a necessary step involves subjecting them to image preprocessing. This process encompasses various adjustments, such as modifying size, orientation, and color. The goal of preprocessing is to enhance the quality of the image stacks, thereby enabling more effective analysis. By eliminating undesired distortions and enhancing specific qualities, preprocessing ensures that the image stacks align with the requirements for the ImageJ plugins (HWada TPIV and TrackMate). Consequently, image preprocessing is essential to ensure proper functioning of the software and the attainment of desired results.

For our purpose, the objective is to eliminate the background entirely and produce binary frames that exclusively contain visible bacteria. To minimize errors in PIV and PTV analysis, it is essential to supply the ImageJ plugins with binary or grayscale frames.

After obtaining binary image stacks from videos, 8 bit-color (binary) thresholds are applied to exclude the background. Fortunately, the bacteria tend to appear brighter than the glass substrate in the background, making it possible to effectively exclude most of the background in this manner. However, in cases where uneven microscope lighting causes uneven illumination, a Fast Fourier bandpass filter is applied to achieve uniform contrast throughout the images. This significantly improves the thresholding process. Background elements that could not be excluded, typically manifest as small irregularly shaped dots. To remove them, the "Analyze Particles" tool in ImageJ is utilized, and a mask is set to eliminate dots of specific sizes and shapes. For spots that are similar in size to bacteria, they need to be either manually eliminated or ignored during the PIV and PTV analysis. Finally, to enhance the detection capabilities of the HWada TPIV and TrackMate plugins, the bacteria are dilated, and their borders are smoothed.

### 1.2 HWada TPIV Data Export

The ImageJ plugin utilized in this process imports videos as image stacks and performs PIV analysis on every pair of images.[1,2] For instance, if an image stack with N images is imported, the plugin generates N-1 PIV results for each image pair (e.g., Pairs: AB, BC, CD from images: A, B, C, D, etc. respectively). The adjustable parameter of interest is the interrogation window size, which is set to 8x8 in this case. The plugin outputs three separate delimited text files: one containing the coordinates of grid points (nodes), another containing the magnitude of velocity at these nodes, and the third containing the angle of velocity at those nodes.

### 1.3 TrackMate Data Export

TrackMate imports image stacks and performs tracking of individual bacteria.[3] However, tracking can encounter challenges when multiple bacteria intersect or cluster together. To address these situations, a combination of the Overlap tracker and Kalman filter tracking is utilized. The overlap tracker excels at distinguishing particles during overlaps, while the Kalman filter leverages past direction and velocity information to predict the bacteria's future trajectory. The plugin generates two CSV files: one for the Kalman filter and another for the overlap filter. These files contain various columns with data such as bacteria IDs, positions over time, and velocity over time etc.

The outputs provide valuable information about the tracked bacteria, aiding in further analysis and understanding of their movement patterns.

**1.4 Python PIV and PTV Analyses**

PIV is a useful method to understand flow direction in a system as it calculates in which directions the pixels in an ordered image stack (video) are shifting. By removing the background, and only providing it bacteria positions in images, the PIV analysis pixel shifts can be directly translated into the bacteria velocity direction from $t = n^{th}$ frame to $t + \Delta t = (n + 1)^{th}$ frame. In the PIV analysis, the three text files containing node coordinates, magnitudes, and angles for image pairs are imported. By taking the sum of all the PIV frames we will find the average direction of the pixel shifts (flow direction) at each node and dividing by the number of image pairs, we obtain the average velocities of each bacterium for the entire length of the video. The angle distribution is subsequently determined for each node within the 8x8 interrogation window. Nodes with very small average velocities, which often indicate errors or regions where the background was not properly excluded, are filtered out.

In the PTV analysis, the CSV files are imported, and only the columns for Track ID and positions at each time are considered. From these columns, the angle distribution is calculated for each bacterium in every pair frame. Finally, the distributions for all pair frames are summed and divided by the number of frames. This process is performed for both the overlap tracking data and the Kalman filter data, and the averages of the two distributions are obtained.

Finally, we normalize the PIV data by using the number of nodes, and PTV data using the number of bacteria tracks and generate figures for the orientational distribution. These analyses provide insights into the angle distributions for both PIV and PTV, allowing for a comprehensive understanding of the motion patterns and behaviors of the studied bacteria.

**1.5 Orientation J Analysis**

An ImageJ plugin developed by the Biomedical Imaging Group known as Orientation J was employed to determine the positional orientation of the swimmers.[4] Within this plugin, the user specifies the dimensions of a Gaussian-shaped window, and the software calculates the structure tensor for each pixel within the image by systematically applying the Gaussian analysis window across the entire image. The local orientation characteristics are then computed and represented as color images, typically encoding orientation information through color hues. The orientation is assessed for each pixel in the image based on the structure tensor.[5] Subsequently, a histogram of orientations is constructed, considering pixels with a coherency greater than the specified minimum coherency and an energy level exceeding the minimum energy threshold. This histogram is a weighted histogram, with the weight assigned based on the coherency value itself. The minimum coherency is expressed as a percentage, as the coherency factor falls within the range of 0 to 1. Similarly, the minimum energy is expressed as a percentage of the maximum energy present in the image. For our specific purpose, we set the window size to 5 pixels, coherency threshold to 0% and the minimum energy threshold to 50%. The output of the analysis gives a list of angles from -90° (270°) to +90° in polar coordinates with the number of pixels containing bacteria (bacteria count) in the binary image associated with these angles.

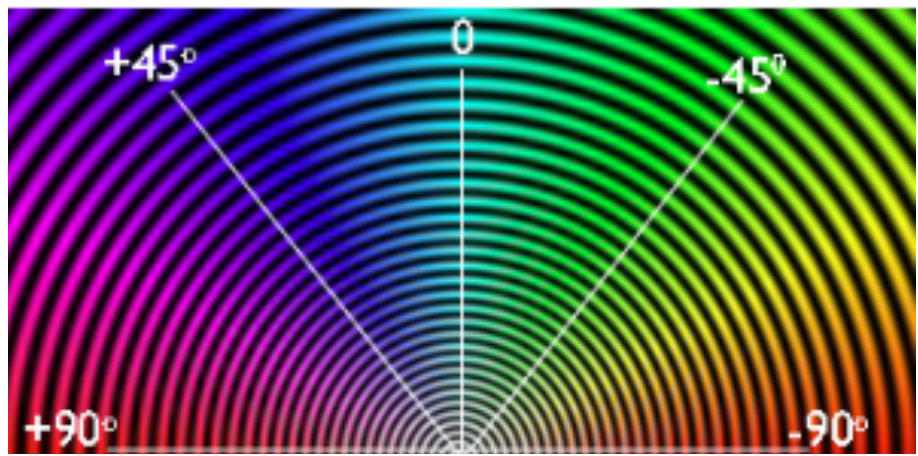

**Figure S1:** Circular color map coding.

## 1.6 Summary of number of samples and number of bacteria studied.

|  |  | h=0 | | h=10 | |
|---|---|---|---|---|---|
|  |  | Number of samples | Number of bacteria | Number of samples | Number of bacteria |
| No LC film | No rubbing | 6 | 695 | 5 | 346 |
|  | Rubbing at 180° | 6 | 649 | 10 | 4236 |
| No *E* field | No rubbing | 12 | 1301 | 10 | 2350 |
|  | Rubbing at 180° | 13 | 3883 | 9 | 2000 |
| *E* field along 270° | No rubbing | 8 | 2060 | 5 | 1473 |
|  | Rubbing at 180° | 8 | 3024 | 5 | 1980 |
| Rubbing along 90° | *E* field at 90° | 10 | 3357 | 10 | 3107 |
|  | *E* field at 270° | 12 | 1960 | 11 | 2165 |

**Table S1.** Summary of the number of samples and number of bacteria studied in the different configurations. The purpose of using several samples in each configuration is to minimize the thickness gradient of the Terrific Broth that could cause flow of the TB and the bacteria inside.

## 2. Additional Results

### 2.1 PTV Results at h=10

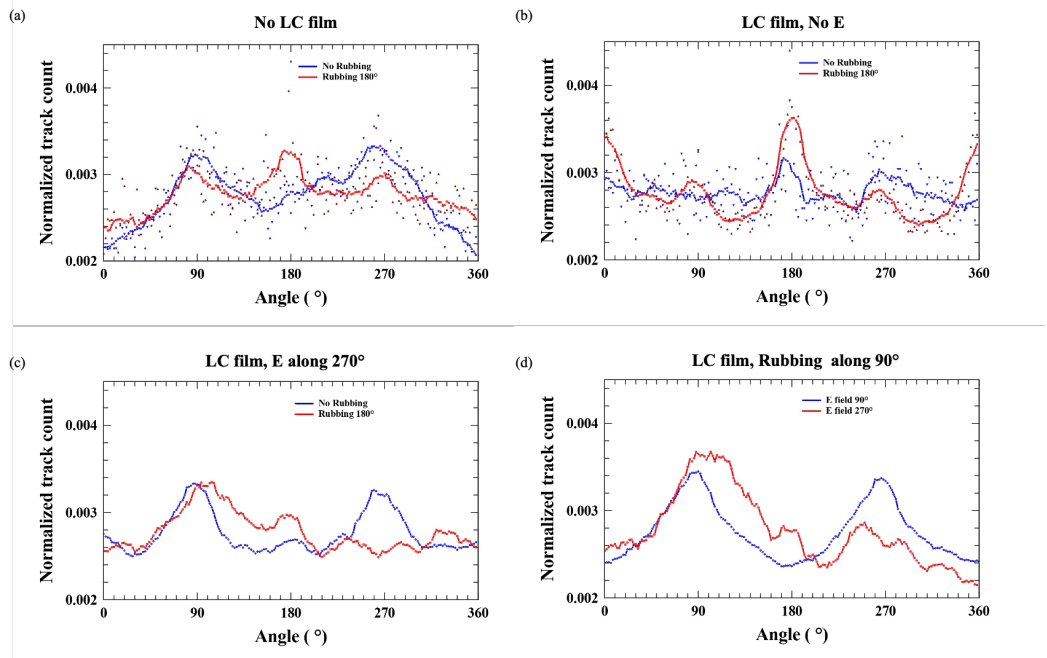

**Figure S2:** PTV results in *h*=10 range.

## 2.2 PIV results

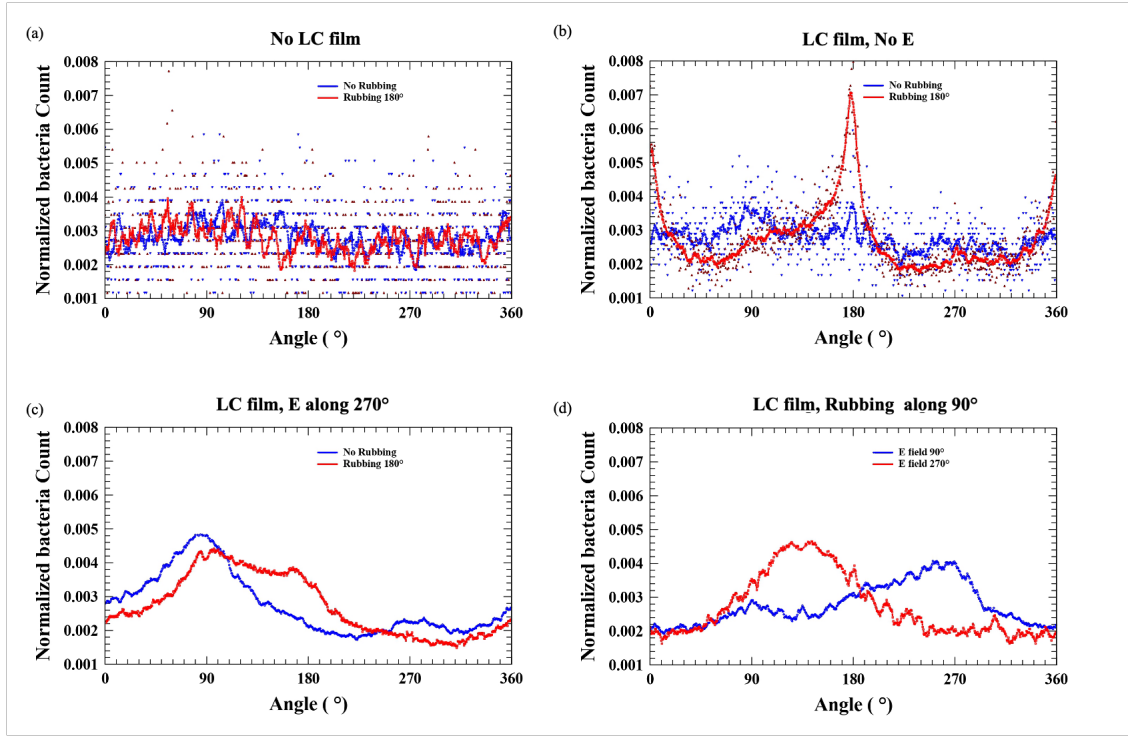

**Figure S3:** PIV results in *h*=0 range.

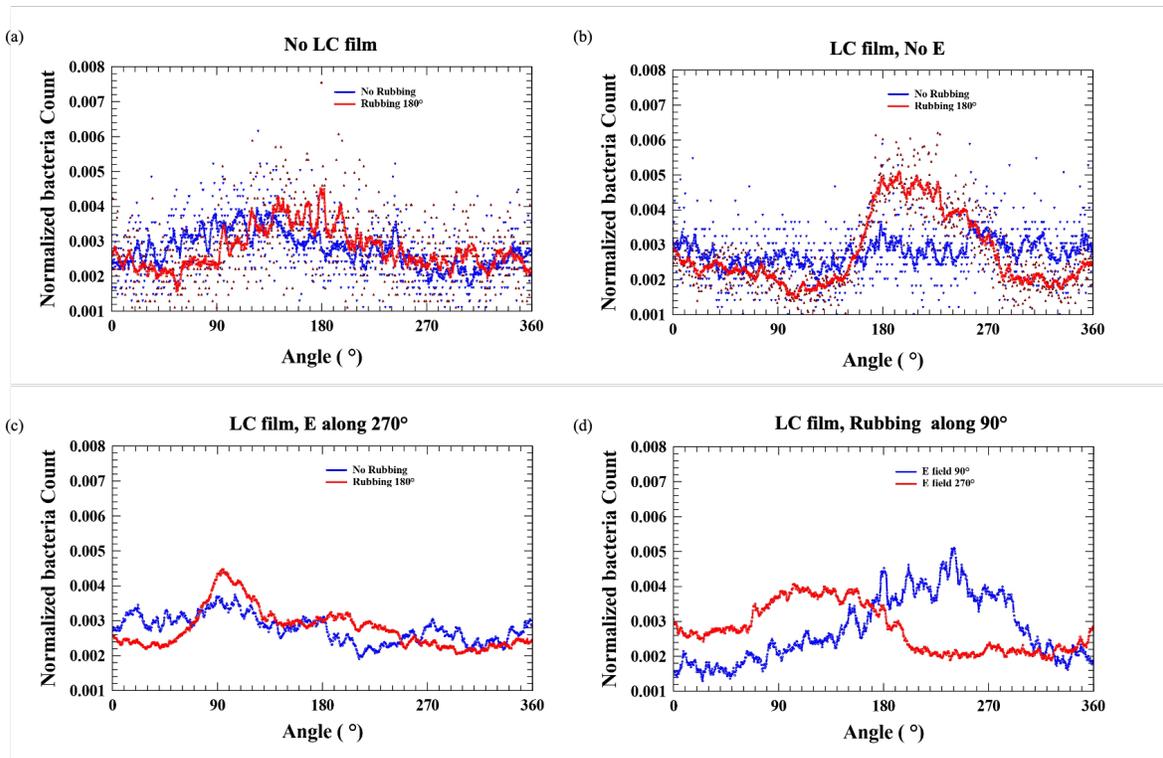

**Figure S4**: PIV results in *h*=10 range.

## 2.3 Orientation J Results

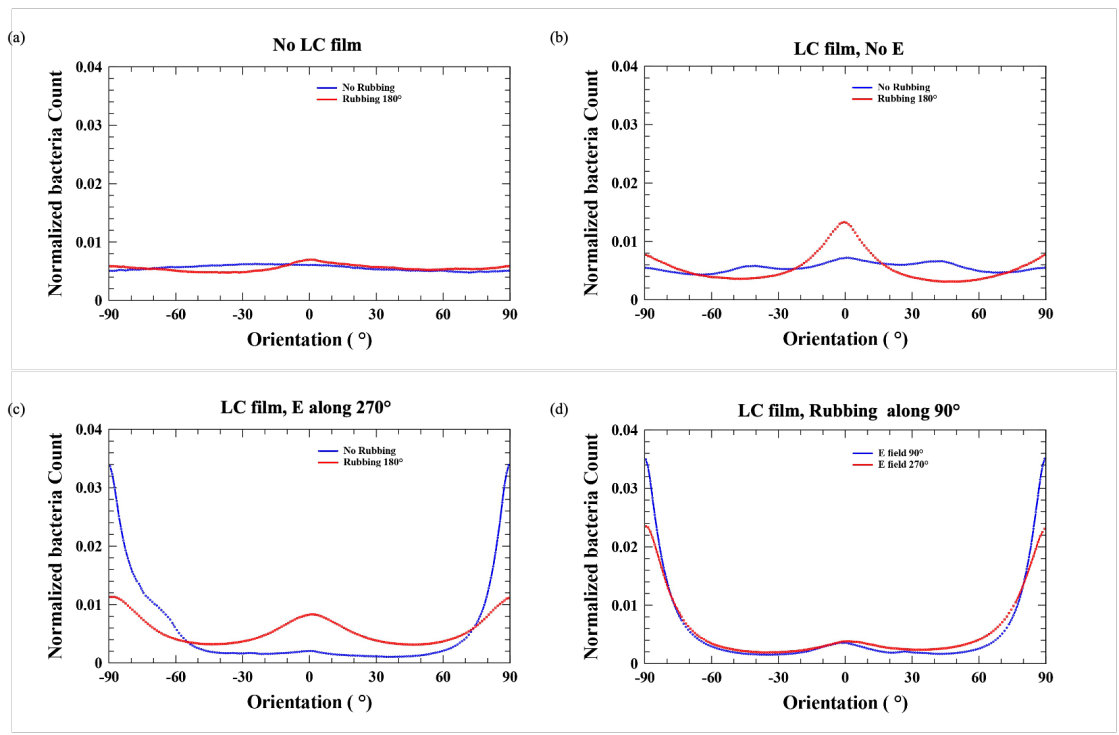

**Figure S5:** Orientation J results *h*=0 in range.

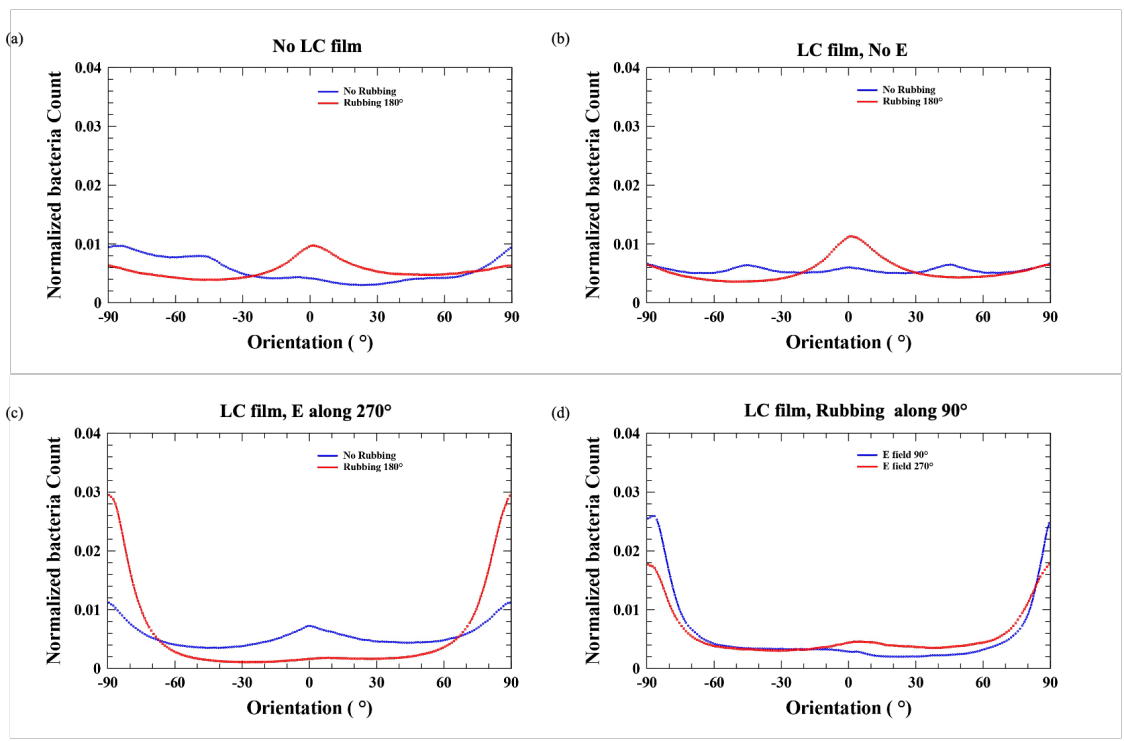

**Figure S6**: Orientation J results in *h*=10 range.

## 3. Statistics

### 3.1 Trajectory Statistics Method:

Subsequent analysis involved fitting trajectory data for each bacterium with a minimum trajectory length of 200 pixels (80 µm) along the trajectory direction to a two-dimensional line employing the method of least squares (see Figure S7), which involves minimizing the sum of squared deviations from the fitted line - enabling a quantitative estimation of fit parameter for the following equation, $F(x) = ax + b$, where $a, b$ are the fit parameters. Following the fitting process, an R-square test was conducted to assess the goodness of fit, measuring the proportion of variance in the dependent variable explained by the independent variable in the linear regression model.[6] After careful visual inspection, a threshold was set for identifying linear trajectories based on an R-square test score exceeding 0.8 (with a maximum score of 1.0) - which are trajectories with arclength of 200 pixels (80 µm) and radius of curvature greater than approximately 316 pixels (126.4 µm).

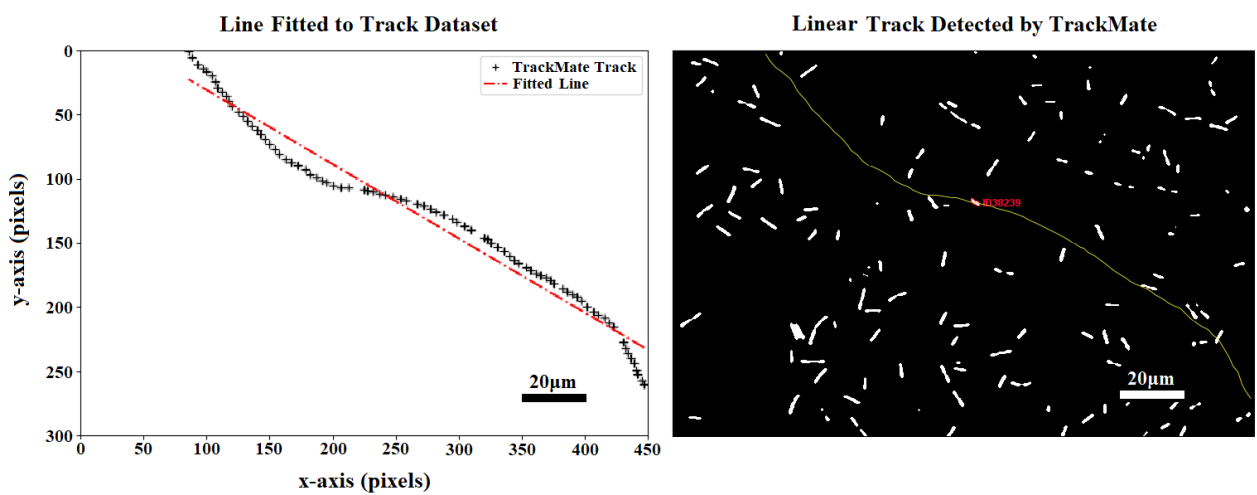

**Figure S7:** Least square fitting of a two-dimensional line to bacteria track coordinates. (Left): Equation of line fitted with calculated parameters to the swimmer track. The $R^2$ test score for this track was $0.97 > 0.8$ (Linear Threshold). (Right): Snapshot of 1 frame from a video where TrackMate was used to find the full track of one active swimmer.

Trajectories were then fitted to an ellipse (see Figure S8) using the method of direct linear least square fitting of ellipses demonstrated by Fitzgibbon et al.[7]

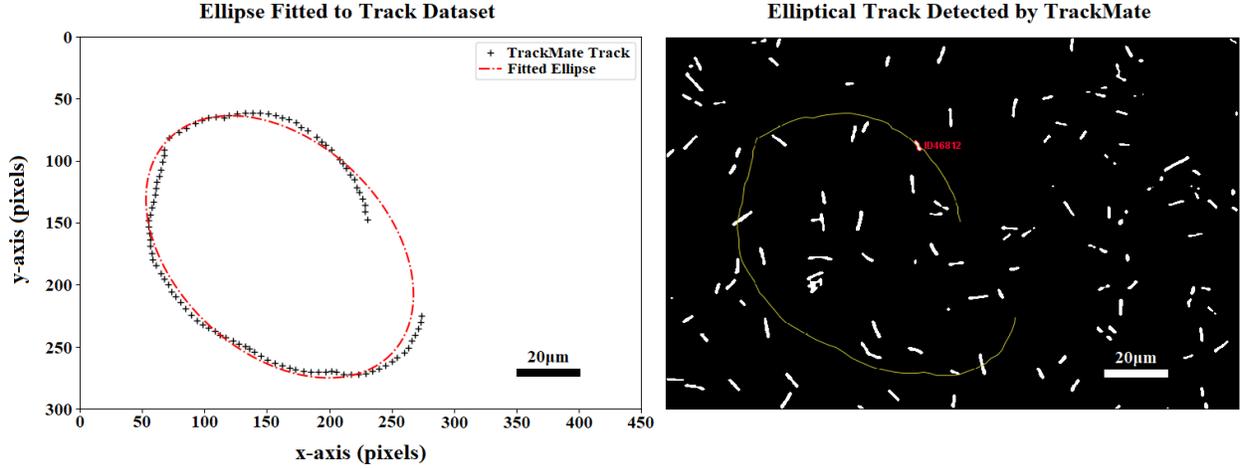

**Figure S8:** Direct least square fitting of ellipse to bacteria track coordinates. (Right) Snapshot of 1 frame from a video where TrackMate was used to find the full track of one active swimmer. (Left) Equation of an ellipse fitted with calculated parameters to the swimmer track.

In image recognition and analysis, it's quite common to encounter the task of fitting a set of data points to an ellipse on the $xy$ plane. Interestingly, this problem can often be tackled using a linear least squares solution, primarily because the general equation of any conic section can be formulated as follows:

$$F(x,y) = ax^2 + bxy + cy^2 + dx + ey + f = 0$$

which is linear in its parameters $\gamma = [a, b, c, d, e, f]$. In Fitzgibbon et al[7] they used the fact that the parameter vector $\gamma$ can be scaled arbitrarily to impose the equality constraint $4ac - b^2 = 1$, thus ensuring that $F(x,y)$ is an ellipse. The least-squares fitting problem can then be expressed as minimizing $||D\gamma||^2$ subject to the constraint $\gamma^T C\gamma = 1$, where the design matrix:

$$D = \begin{pmatrix} x_1^2 & x_1 y_1 & y_1^2 & x_1 & y_1 & 1 \\ \vdots & \vdots & \vdots & \vdots & \vdots & \vdots \\ x_n^2 & x_n y_n & y_n^2 & x_n & y_n & 1 \end{pmatrix}$$

represents the minimization of $F(x,y)$. The method of Lagrange multipliers, as introduced by Gander[8], yields the conditions: $S\gamma = \lambda C\gamma$ and $\gamma^T C\gamma = 1$, where the scatter matrix, $S = D^T D$. There are up to six real solutions, $(\lambda_j, \gamma_j)$ and, it was claimed, the one with the smallest positive eigenvalue, $\lambda_k$ and its corresponding eigenvector, $\gamma_k$, represent the best fit ellipse in the least squares sense. Fitzgibbon et al[7] initially introduced an algorithm in MATLAB that implemented this method, later refined by Halíř and Flusser[9] to enhance its reliability and numerical stability. It is this improved algorithm that was used to find fit parameters for elliptical bacterium trajectories. Moreover, the algorithm possesses an additional capability to discard datasets that do not conform to elliptical patterns if they fail to satisfy the Gander conditions.[7]

The tracks dismissed by both fitting procedures constitute around 10% of the overall combined count of bacteria tracks across all videos. These disregarded tracks exhibit irregular shapes as shown in Figure S9, likely remnants of artifacts resulting from either bacterium tracks suddenly ending due to it swimming to different focal plane (movement in $z$-axis, not in $xy$-axis) or errors in Advanced Kalman Tracking algorithm from TrackMate. Consequently, these rejected tracks are omitted from the trajectory statistics.

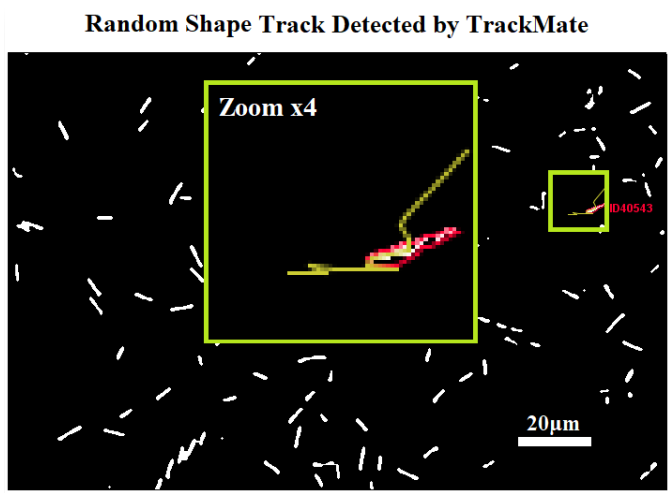

**Figure S9:** Non-linear and non-elliptical trajectory shapes detected by TrackMate. The bacteria in the image swims to a different focal plane, resulting in a short track that is rejected by both linear and elliptical fitting routines.

### 3.2 Trajectory Statistics Results

Figure S10 illustrates that in the absence of liquid crystal, the ratio of linear to elliptical track count is notably elevated. The aggregated data indicates that trajectories of bacteria swimming closest to the surface of glass substrate exhibit a slightly larger circularity (ratio of minor to major axes of elliptical trajectories) compared to those swimming at higher planes. Considering the dependence on the surface treatment there is no clear trend.

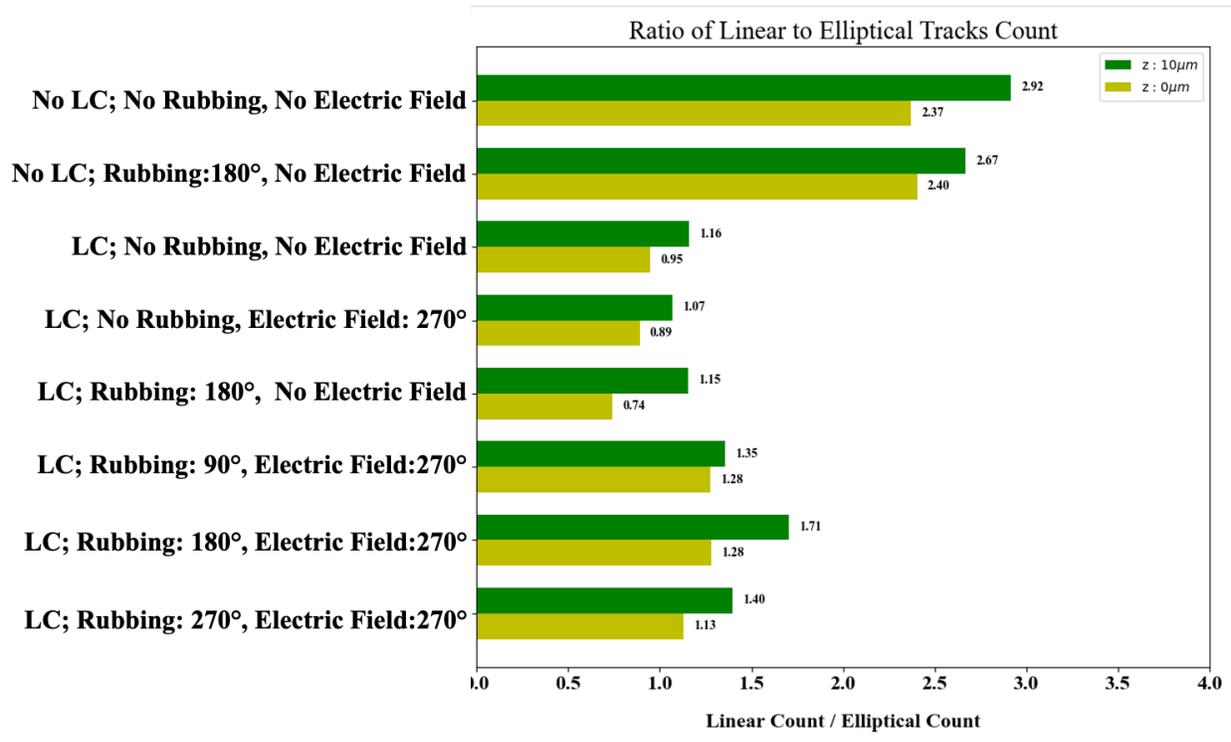

**Figure S10:** Ratio of linear and elliptical swimming trajectories of bacteria under different testing conditions (i.e. variable rubbing, liquid crystal shearing and electric field direction). Quadratic and elliptical track counts were added together and compared to linear track counts for the ratio.

**References**

[1]    J. Schindelin, I. Arganda-Carreras, E. Frise, V. Kaynig, M. Longair, T. Pietzsch, S. Preibisch, C. Rueden, S. Saalfeld, B. Schmid, J. Y. Tinevez, D. J. White, V. Hartenstein, K. Eliceiri, P. Tomancak, A. Cardona, *Nat Methods* **2012**, *9*, 676.

[2]    C. A. Schneider, W. S. Rasband, K. W. Eliceiri, *Nat Methods* **2012**, *9*, 671.

[3]    D. Ershov, M. S. Phan, S. U. Pylvänäinen, J.W. Rigaud, L. Le Blanc, J. R. Charles-Orszag, A., Conway, R. F. Laine, N. H. Roy, D. Bonazzi, G. Duménil, *Bioarxiv* **2012**, 2021.

[4] P. Zsuzsanna, S. Martin, S. Daniel, U. Michael, in *Adv Anat Embryol Cell Biol*, **2016**, pp. 69–94.

[5] R. Rezakhaniha, A. Agianniotis, J. T. C. Schrauwen, A. Griffa, D. Sage, C. V. C. Bouten, F. N. Van De Vosse, M. Unser, N. Stergiopulos, *Biomech Model Mechanobiol* **2012**, *11*, 461.

[6] D. Chicco, M. J. Warrens, G. Jurman, *PeerJ Comput Sci* **2021**, *7*, e623.

[7] A. Fitzgibbon, M. Pilu, R. B. Fisher, *IEEE Trans Pattern Anal Mach Intell* **1999**, *21*, 476.

[8] W. Gander, *Numer Math (Heidelb)* **1980**, *36*, 291.

[9] R. Halíř, J. Flusser, *Proc. 6th International Conference in Central Europe on Computer Graphics and Visualization. WSCG* **1998**, *98*, 125.